\documentclass[%
 reprint,
superscriptaddress,
 amsmath,amssymb,
 aps,
]{revtex4-1}

\usepackage{graphicx}
\usepackage{dcolumn}
\usepackage{bm}

\begin{document}


\title{Tailoring the opto-electronic response of graphene nanoflakes by size and shape optimization}

\author{Raquel Esteban-Puyuelo}
 \affiliation{Division of Materials Theory, Department of Physics and Astronomy, Uppsala University, Box-516, SE 75120, Sweden}

\author{Rajat Kumar Sonkar}%
\affiliation{Centre for Modeling and Simulation, Savitribai Phule Pune University, Ganeshkhind, Pune 411007, India}

\author{Bhalchandra Pujari}
\affiliation{Centre for Modeling and Simulation, Savitribai Phule Pune University, Ganeshkhind, Pune 411007, India}

\author{Oscar Gr\aa n\"as}
\email{Corresponding Author: oscar.granas@physics.uu.se}
\affiliation{Division of Materials Theory, Department of Physics and Astronomy, Uppsala University, Box-516, SE 75120, Sweden}

\author{Biplab Sanyal}
\email{biplab.sanyal@physics.uu.se}
\affiliation{Division of Materials Theory, Department of Physics and Astronomy, Uppsala University, Box-516, SE 75120, Sweden}

\footnotetext{$\ast$~Electronic Supplementary Information (ESI) available: [details of any supplementary information available should be included here]. See DOI: 00.0000/00000000.}

\date{\today}

\begin{abstract}
The long spin-diffusion length, spin-lifetimes and excellent optical absorption coefficient of graphene provide an excellent platform for building opto-electronic devices as well as spin-based logic in a nanometer regime. In this study, by employing density functional theory and its time-dependent version, we provide a detailed analysis of how the size and shape of graphene nanoflakes can be used to alter their magnetic structure and optical properties. As the edges of zigzag graphene nanoribbons are known to align anti-ferromagnetically and armchair nanoribbons are typically non-magnetic, a combination of both in a nanoflake geometry can be used to optimize the ground-state magnetic structure and tailor the exchange coupling decisive for ferro- or anti-ferromagnetic edge magnetism, thereby offering the possibility to optimize the external fields needed to switch magnetic ordering. Most importantly, we show that the magnetic state  alters the optical response of the flake leading to the possibility of opto-spintronic applications.
\end{abstract}

\maketitle


\section{Introduction}

Since the discovery of graphene in 2004 \cite{Novoselov2000}, numerous studies have revealed its extraordinary electronic \cite{Bolotin2008}, thermal \cite{Balandin2008}, mechanical \cite{Lee2008} and even superconducting \cite{Cao2018} properties, to name a few. This makes graphene one of the most promising materials for energy efficient devices. However, being a semi-metal, graphene has limited scope of applications in transistor-based technology as one needs a large on-off ratio for transistor operations. For this purpose, several routes have been considered, viz., chemical functionalization \cite{Yan2010} and realizing nanostructures in the form of nanoribbons \cite{Son2006, Son2006a} and nanoflakes \cite{WeiL.Wang2007, Silva2010}. The advantage of these nanostructures is the control of properties via their size, shape and edge structures. Specifically, the quantum confinement in nanoflakes gives the possibility to tune the gap and optical properties as a function of size and shape. Moreover, electronic and magnetic properties are crucially dependent on the zigzag and armchair edge structures. 

\begin{figure}[ht]
\includegraphics[width=\linewidth]{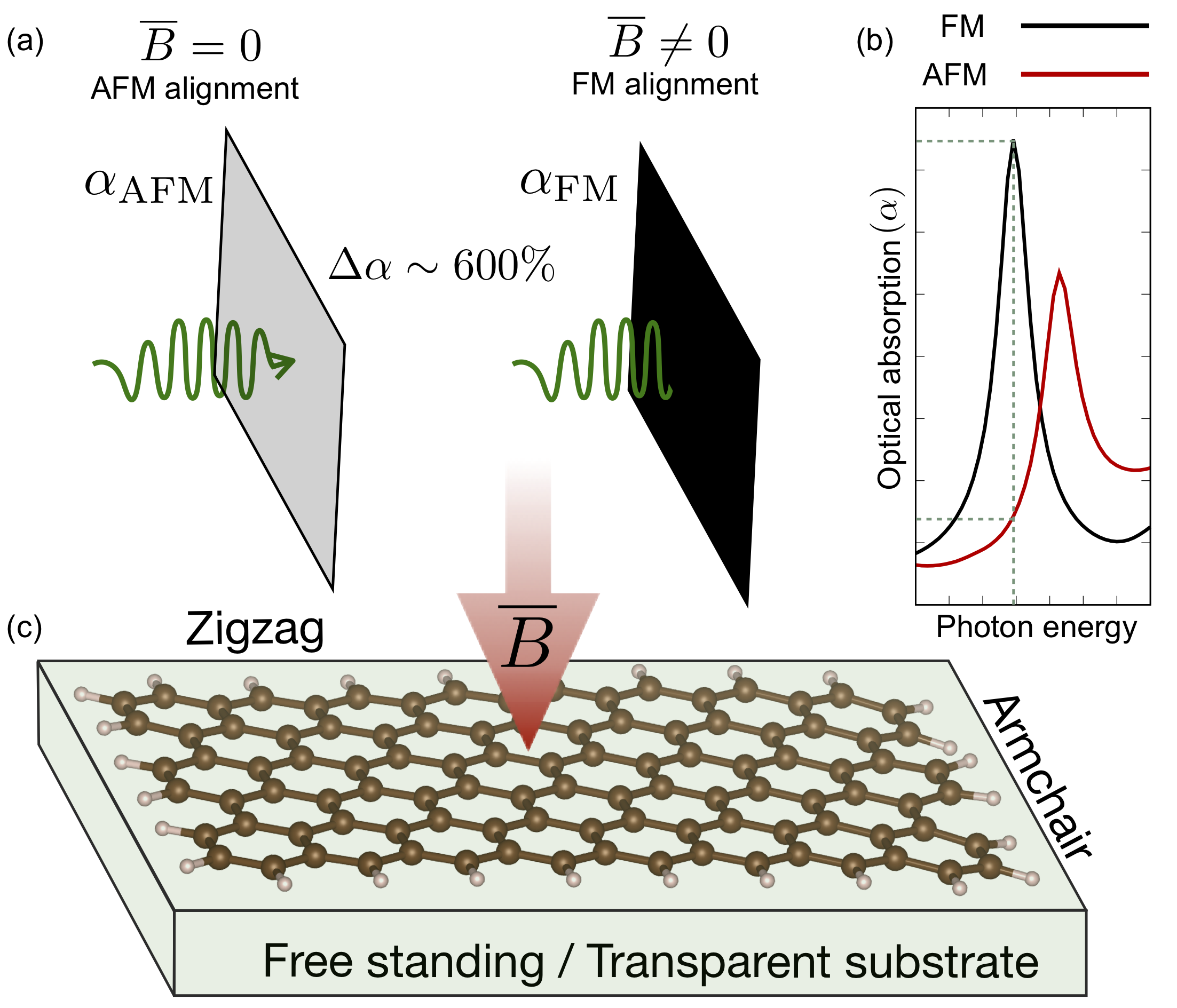}
\caption{Schematic view of an opto-electronic device based on rectangular graphene nanoflakes. The size and shape affect the exchange coupling between the edges. Hence, they can be tailored to optimize the sensitivity to magnetic fields, which in turn alters the opacity of the flake. The top left section (a) shows the change in opacity when going from an antiferro- to a ferromagnetic alignment, the top right segment (b) shows the change in the optical absorption spectrum around a relevant energy. The bottom segment (c) shows a rectangular flake with armchair and zigzag edges.}
\label{fig: figure1}
\end{figure}

The remarkable optical absorption characteristics of graphene in a wide range of frequency and its extraordinarily high mobility offers a great potential in developing high speed and flexible opto-electronic devices. The dielectric properties of graphene are known to be gate-tunable \cite{Santos2013}, indicating a strong coupling between external electric field and the opto-electronic response. We show that a similar strong coupling exists between the dielectric properties and external magnetic fields in nanoflakes. It should be noted that the energy  consumption of electronic devices is a major issue for information and communications technology \cite{Miller2017}. Here, we address the issue of energy-efficiency by showing that the opto-electronic response can be optimized by the shape and size of graphene nanoflakes. The electronic structure in the proposed setting is tuned to have an exchange coupling that allows for energy-efficient switching between ferromagnetic (FM) and anti-ferromagnetic (AFM) coupling across the zigzag edges. The magnetic state of the flake has a clear signature in the optical transmission. For specific frequencies, the opacity of the flake can be switched using small magnetic fields, as represented in figure \ref{fig: figure1}.

There are currently several methods to fabricate graphene nanoribbons, the first ones being chemical \cite{Li2008} or top-bottom approaches such as unzipping them from carbon nanotubes \cite{Jiao2009, Kosynkin2009}. Achieving high accuracy in the edge termination, width and length of the nanoribbon is very difficult via these routes, but nowadays the bottom-up approach \cite{Cai2010, Ruffieux2016} has become standard because it allows for atomic precision and control of the width, edge type and termination of the graphene nanoribbons. Rectangular nanoflakes can be obtained with this method by stopping the growth at the desired length. On the other side, hexagonal graphene flakes have extensively been studied experimentally \cite{Dotz2000, Zhi2008}. Moreover, magnetism has been observed in graphene quantum dots experimentally through spin-polarized edge states \cite{Sun2017}.

Inspired by the precise control over size and shape provided by the bottom-up approach, we study ground-state properties of graphene nanoflakes of varying size and shape, including magnetic ordering across their edges. To this end, we use first-principles density-functional theory (DFT). Further, we investigate the optical properties using time-dependent density functional theory (TD-DFT) \cite{Marques2004, Marques2006}, which allows for the study of the response of a system under the application of a time-dependent external field. 

In contrast to previous studies of the shape and size dependence of the optical properties of graphene nanoflakes, and most prominently, the TD-DFT in conjunction with density functional tight-binding study by Wettstein et al.\cite{MansillaWettstein2016}, we focus on the coupling between the optical properties and magnetic ordering. We show that the tuning of the magnetic interactions is possible for geometries of rectangular type, but not for hexagonal. For benchmark purpose, we provide in addition, details of the impact of specific implementations on the calculated optical properties, including hexagonal flakes and smaller molecules in the electronic supplementary information (ESI).

The structure of the paper is as follows. In the next section, we present the details of the computational part including the methodology of the calculation of optical absorption spectra using the TDAP \cite{Meng2008, Kolesov2016} method based on the SIESTA \cite{Soler2002} code. Thereafter, we present the results obtained for magnetic ground states and optical properties. Finally, we show the size and shape dependence on optical spectra of zigzag and armchair nanoflakes followed by conclusions. In the ESI, we present additional computational details, information about auxiliary geometries and a benchmark of the optical properties of chosen small molecules to show that the methodology used is able to capture important features in the absorption spectrum.

\section{Methodology}
We use spin-dependent density functional theory as implemented in the SIESTA package to study the magnetic ordering and its dependence on the number of armchair vs zigzag edges that are present in nanoflakes. The SIESTA code uses a local basis comprising a set of numerical atomic orbitals built from  norm-conserving pseudopotentials based on the Kleynman-Bylander scheme \cite{Kleinman:1982cx}. Apart from SIESTA, we have also used Octopus and NWChem codes to benchmark small sized carbon based molecules. The details are given in the ESI.

The structures have been relaxed in order to minimize the forces down to 0.005 eV/\AA\ using a split triple zeta with polarization (TZDP) basis with a mesh cutoff of 120 Ry and an energy threshold of 10$^{-4}$ eV. The electronic exchange and correlation is treated with the functional by Perdew, Burke and Ernzerhof. The small flakes and molecules displayed in the ESI are placed in a cubic box of 20 \AA\ side, whereas the large rectangular flakes are in a box of 40 \AA\ size.

The optical properties are studied using a real-time version of time-dependent density functional theory (RT-TDDFT) implemented on top of the SIESTA package. An initial state is propagated in time according to the time dependent Kohn-Sham (TD-KS) equations:
\begin{eqnarray}
i\hslash\frac{\partial \phi_l(\bm{r},t)}{\partial t}=\mathcal{H}_{\mathrm{KS}}[n(\mathbf{r}),m(\mathbf{r})]\phi_l(\bm{r},t),
\label{eq: TD-KS}
\end{eqnarray}

where $\phi_l$ are the single electron KS wave functions and $n(\mathbf{r})$ and $m(\mathbf{r})$ are the charge and magnetization densities. The $\phi_l$'s in turn are expressed in basis functions $\chi$, which are atomic-orbitals carrying information about the locality of the wave-function in real and spin-space.

The dependence of $\mathcal{H}_{\mathrm{KS}}$ in the charge and magnetization density allows us to extract the response of the nanoflakes for different magnetic structures.

The dielectric properties of our systems are extracted by studying the dynamical response to the perturbation created by an external electric field. We monitor the evolution of the electric dipole moments, calculated through
\begin{eqnarray}
p_{j,k}(t)=\mathrm{Tr}\left[\bm{D}^k\rho(t)\right],
\end{eqnarray}
where $j$ and $k$ indicate the directions of the applied field and measurement respectively and $\rho(t)$ is the time-dependent density matrix expressed in the basis functions $\chi$. The transition dipole tensor operator is defined as
\begin{eqnarray}
\bm{D}^k_{\mu\nu}=\left\langle \chi_{\mu} \left| \hat{e}_k \cdot \mathbf{r}  \right| \chi_{\nu}\right\rangle,
\end{eqnarray}
for direction $\hat{e}_{k}$. The density matrix $\rho(t)$ carries information about the magnetism in the system. In the current work, we have considered collinear magnetic structures. $\bm{D}$ is related to the polarizability tensor $\alpha(t)$ by
\begin{eqnarray}
p_{j,k}(t)=\int_{-\infty}^t \alpha_{j,k}(t-t')E_j(t')dt',
\end{eqnarray}
where $E_j$ is the $j$th component of the external electric field. In our case, we study the impulse response by applying a kick-pulse
\begin{eqnarray}
E(t)=E_0\delta(t-t_0)\hat{e}_j,
\end{eqnarray}
which after Fourier transform yields
\begin{eqnarray}
\alpha_{j,k}(\omega)=\frac{p_{j,k}(\omega)}{E_0}.
\end{eqnarray}
The absorption cross section can be obtained via
\begin{eqnarray}
\sigma_{j,k}(\omega)=\frac{4\pi\omega}{c}\mathrm{Im}\left[\alpha_{j,k}(\omega)\right].
\label{eq: 1/3 trace}
\end{eqnarray}


The Verlet algorithm is used for the  time evolution, where 10000 steps of 24.18 attoseconds have been simulated. The step size is small enough so that the integration algorithm is stable but large enough to achieve a long simulation time with a reasonable computational cost. This gives a total simulation time of 242 fs and is equivalent to a precision of 0.017 eV according to the sampling theorem. For each structure, three separate time evolution runs have been performed by applying an electric field in the x, y or z direction (the flakes are on the x-y plane). Then, the dipole moment has been calculated for each of the runs and the optical spectrum has been obtained by summing up the trace of the absorption cross-section as shown in eq \ref{eq: 1/3 trace}. The spectra have been normalised to the number of valence electrons for each system.

The focus of this study lies in the study of rectangular graphene nanoflakes (RGNF), and we denote them by adapting the standard convention used for graphene nanoribbons \cite{Fujita1996,Son2006}. RGNF are classified by the number of horizontal zigzag chains $N_z$ and vertical armchair dimer lines $N_a$, and hence, from now on will be denoted as $N_a\times N_z$-RGNF. RGNF have parallel zigzag edges in the horizontal direction and parallel armchair edges in the vertical direction.

The benchmark provided in the ESI, performed for smaller molecules and flakes of different shapes, indicates that while the different implementations show some differences (see ESI figure 4-5), trends are always similar and relative peak positions in the optical absorption spectra are conserved.

\section{Results and discussion}

\begin{figure}[ht]
\includegraphics[width=\linewidth]{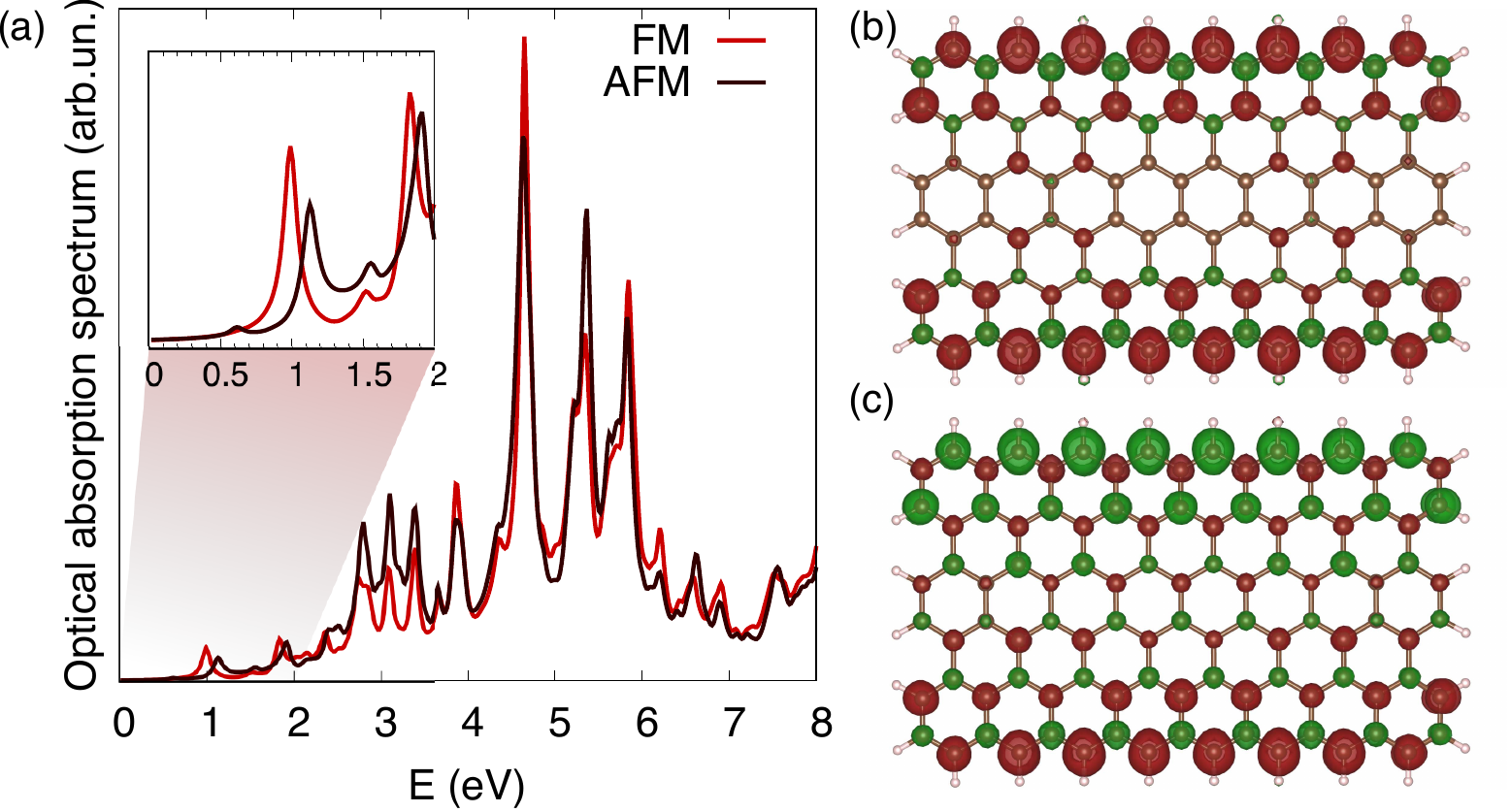}
\caption{\label{fig: figure2} (a) Optical absorption spectra and magnetization densities for (b) ferromagnetically and (c) antiferromagnetically coupled edges in the 17$\times$6-RGNF. The red (green) color represents positive (negative) magnetization density with an isovalue of 0.002 $\mu_B/$\AA$^3$. In the inset of (a), an expanded view of the optical absorption spectra at lower energies is shown.}
\end{figure}

Zigzag edges in graphene nanoribbons are antiferromagnetically coupled, while armchair edges are non-magnetic \cite{Son2006}. As previously mentioned, both zigzag and armchair edges are simultaneously represented in RGNFs. As we show in figure \ref{fig: figure2}, there is a clear difference in the absorption spectrum in AFM compared to FM ordering. 
In the following, we will show that the shape and size can be optimized to tailor the switching conditions between the two solutions, and thus also the optical absorption spectrum.

To investigate the conditions of stability for the magnetic ordering, we compare the total-energy between the AFM and FM solutions of RGNFs with dimensions 17x$N_{z}$, with $N_{z}\in\{ 2, 4, 6, 8\}$ with parallel zigzag edges in its longer dimension. All atomic coordinates are relaxed in the respective spin-state. The resulting difference in total energy is displayed in table \ref{tab:exchange coupling}. 

\begin{table}[!h]
\begin{center}
\begin{tabular}{|l|c|c|c|}
\hline
Size & $\Delta E$ & $\Delta E$/Zigzag edge atom & $\Delta E$/atom \\
\hline
17$\times$2 & -133.42 & -16.678 & -2.471 \\
17$\times$4 & -80.11 & -10.014 & -0.871 \\
17$\times$6 & -71.45 & -8.931 & -0.550 \\
17$\times$8 & -16.71 & -2.089 & -0.099 \\
\hline
\end{tabular}
\caption{Energy differences between AFM and FM ordering, $\Delta E = E_{\text{AFM}}-E_{\text{FM}}$ for four different ratios of zigzag to armchair. To facilitate comparison with both GNRs and different kind of flakes, both the total energy and the energy per zigzag edge atom is specified. Energy units are in meV. \label{tab:exchange coupling}}
\end{center}
\end{table}

Assuming that the dynamics of the edge magnetism can be described by a Heisenberg model, the total energy difference corresponds to first order to the exchange coupling $\Delta E = J_{\text{Ex}} = E_{AFM} - E_{FM}$, where the sign of $J$ conforms to the Heisenberg spin-Hamiltonian. 
The energy increase from the application of an external magnetic field (assumed to be homogeneous at these length-scales) is attributed to the Zeeman energy. In order to achieve a switching of the magnetization from AFM to FM ordering the Zeeman energy has to overcome the exchange energy difference between the two phases.
Due to the reduced symmetry of RGNFs with respect to GNRs it is not as straight forward to find a representative unit of measure that facilitates comparison between flakes of different sizes and semi-infinite GNRs. However, using the exchange energy per zigzag edge atom facilitates comparison with GNRs, and is a measure that improves with the size of the flake. The Zeeman energy is vanishing for the AFM state due to symmetry, but not for the FM state. While extrapolating the energy difference between the AFM and FM states is notoriously difficult, knowledge from ZGNR tells us that the exchange coupling decreases monotonically with increasing distance between the zigzag edges \cite{LeePRB2005}. This is consistent with what is found for magnetic impurities and clusters in non-magnetic hosts, i.e. that increasing separation between the magnetic entities leads to vanishingly small exchange coupling at sufficiently large distances, and superparamagnetism sets in \cite{Sato_RMP2010}. This makes the preference for the AFM state increasingly weak, which leads us to believe that the flakes can be tuned from the strongly coupled narrow flakes of 17x2 and 17x4 in tab. 1, to the weak limit in the long range case. Sufficiently narrow ZNGRs have recently been shown to remain ordered up to room temperature while keeping their semiconducting properties to up to 7nm of width \cite{MagdaNat2014}.

The fact that the energy difference between e.g. the 17$\times$6-RGNF with respect to a nanoribbon of similar width is significantly larger \cite{Son2006} indicates that the strength of the exchange coupling between the edges is different for a flake than a ribbon. A semi-infinite zigzag graphene nanoribbon (ZGNR) is known to align in an AFM fashion. The exchange coupling per zigzag unit-cell favours the AFM solution with 4.0, 1.8 and 0.4 meV for the 8-, 16- and 32-ZGNR, respectively \cite{Son2006}. Comparing to the numbers in \ref{tab:exchange coupling}, we observe the same trend of reduced coupling with increased width, but with roughly half the coupling strength. Therefore, it is possible to control the magnetic exchange coupling through the ratio of armchair to zigzag edge atoms, i.e. their size and shape in a relatively easy way for a flake. We have shown unambiguously that the shape of the RGNF allows us to go from a strongly coupled to a weakly coupled to a decoupled system, allowing for energy efficient switching of miniaturised devices. Meanwhile, the flakes are still transparent to low energy excitations due to the finite size, in contrast to ZGNRs.

An important quantity for the optical absorption of the flakes is the single particle eigenvalues of the electronic Hamiltonian. When the edges are ferromagnetically aligned, the exchange splitting induces an energy shift in the order of 0.1-0.5 eV for high lying valence states or low lying conduction band states in the case of FM-17$\times$6-RGNF. For the AFM case, there is an energy-matching of eigenstates associated with eigenfunctions on either side of the flake, hence there is an apparent spin-degeneracy in the eigenvalue spectrum. 
This difference is observed in the small energy region of the optical absorption spectra, also shown in Figure \ref{fig: figure2}: Flake edges with AFM and FM alignments show a 0.2 eV difference in the first peak.

\begin{figure}[ht]
\includegraphics[width=\linewidth]{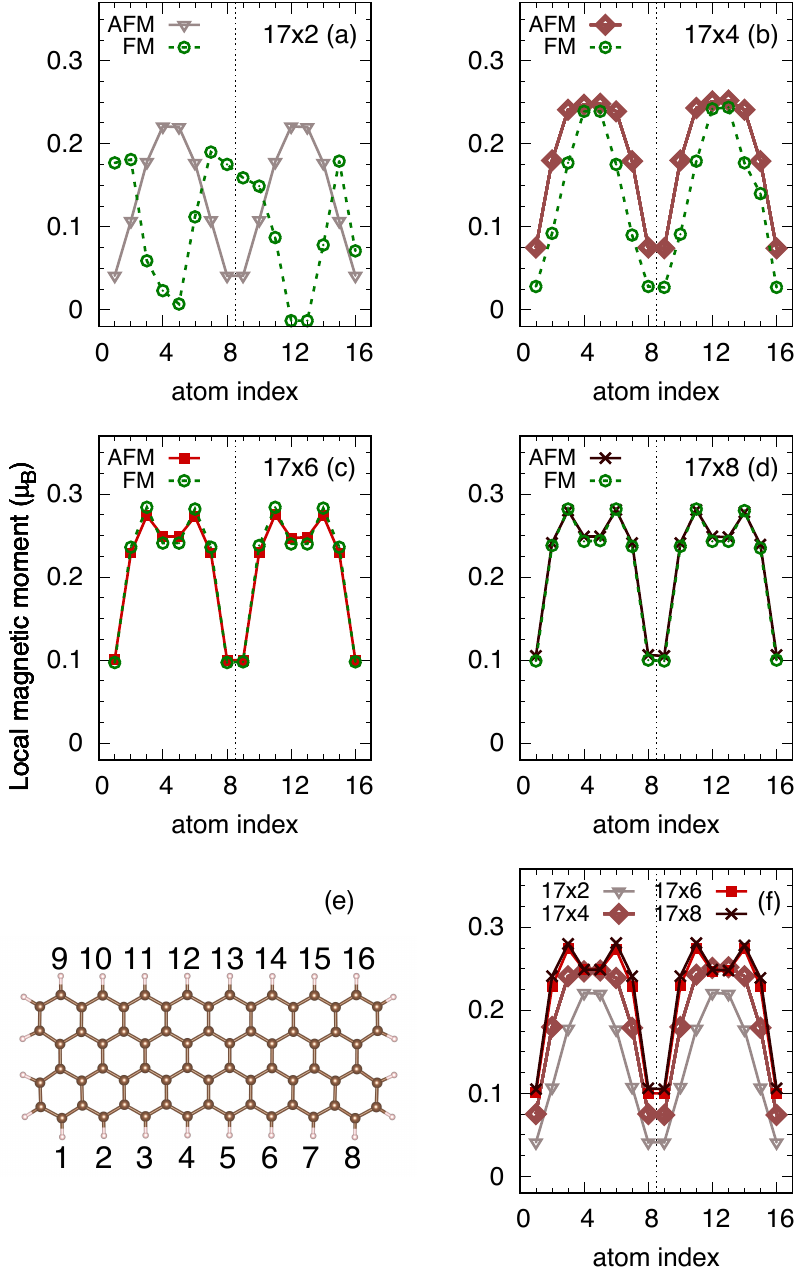}
\caption{\label{fig: figure3} Absolute value of the magnetic moment on the edge atoms of RGNFs with dimensions 17x$N_{z}$, with $N_{z}\in\{ 2, 4, 6, 8\}$. Panel a) represents the 17$\times$2 RGNF, it should be noted that the FM solution is not completely stable, as the center of the zigzag edge (index 5-11) on the RGNF has a different alignment than the corners. However, it is FM aligned between the zigzag edges. It is clear that the more armchair edges we add, separating the zigzag edges further, the more Heisenberg-like the system becomes. Panel e) shows the indexing of the atoms on the zigzag edges. To see the distribution of the magnetization density we refer to the 17$\times$6-RGNF in Figure \ref{fig: figure2}}
\end{figure}

To rationalize the reason behind the apparent sign-change in exchange coupling, we plot the magnetization density iso-surfaces in figure \ref{fig: figure2}. Nearest neighbour carbon atoms are subjected to AFM alignment. When the width (i.e. the armchair dimension) is even in combination with a FM alignment, a frustration is induced in the center rows. For AFM alignment frustration does not occur. However, the armchair edges acquire a magnetic moment, known to be unfavourable for this edge type. The competition of frustration in the center for the FM flake versus magnetism in the armchair edges of the AFM flake has as a consequence that the total energy of the solutions reach energy degeneracy much sooner than for GNRs. Figure \ref{fig: figure3} shows the magnetic moments of the atoms in the zigzag edges for the FM and AFM configurations. Clearly wider flakes are a better approximation to a Heisenberg system, since the magnitude of the magnetic moments are maintained for different magnetic configurations, therefore the energy difference between AFM and FM is a better representation of the exchange interaction mentioned in table \ref{tab:exchange coupling}. Additionally, Figure 2 in the ESI includes the C-C bond lengths along these edges for both AFM and FM  configurations, showing that corner atoms have a shorter bond length, which is translated in a lower magnetic moment.

When the distance between the zigzag edges increases, thermal fluctuations destroy the ordering between the edges already at a low temperature. However, the AFM solution can be stabilized  by increasing the number of zigzag edges, approaching a similar coupling strength to ZGNRs. The possibility to tailor the exchange coupling between the edges allows for making the flake arbitrary susceptible to external magnetic fields. This in turn allows for rapid and energy efficient induction of an FM state for a weakly coupled AFM ground state.

Further, we provide in the ESI, results for rectangular flakes where the symmetry of the edge structure is broken. Due to the broken symmetry of the edge and corner structure, we present energy differences per atom instead of per zigzag edge in this section. Specifically the 16$\times$6-RGNF (armchair edges are closer due to the removal of a vertical dimer line) and 17$\times$5-RGNF (zigzag edges are closer due to the removal of a zigzag line), see ESI figure 1 for details of the structure and magnetization density. Despite the difference in carbon coordination for the corner atoms with respect to the RGNFs presented in the previous section, the same competition between frustration in the center in the FM structure against magnetization in the armchair edges in the AFM structure applies. 
The narrowing of the flake in 17$\times$5-RGNF with respect to 17$\times$6-RGNF leads to an increased stability of the AFM solution that is here favourable by 1.2 meV/atom, larger than the 0.5 meV/atom of the 17$\times$6-RGNF. On the contrary, 16$\times$6-RGNF has the same distance between the zigzag edges as the original flake, but armchair edges are closer, this has little effect on the stability, as the AFM-FM energy difference is 0.37 meV/atom in this case, comparable  to 17$\times$6-RGNF. In addition, we studied flakes in which we removed alternating atoms of the zigzag edge so that the horizontal edge is no longer zigzag, here the modified edge looses the magnetization. This points toward the necessity of zigzag edges for realizing magnetism, similar to what can be expected based on experience with GNR \cite{Son2006}. Furthermore we provide the details of a rhomboid shaped graphene nanoflake with zigzag edges in the ESI. In line with the RGNF results, one of the possible antiferromagnetic configurations is the ground state so that the largest magnetic moments appear on zigzag edges close to the sharp ends perpendicular to the two armchair corners.

\begin{figure*}[ht]
\includegraphics{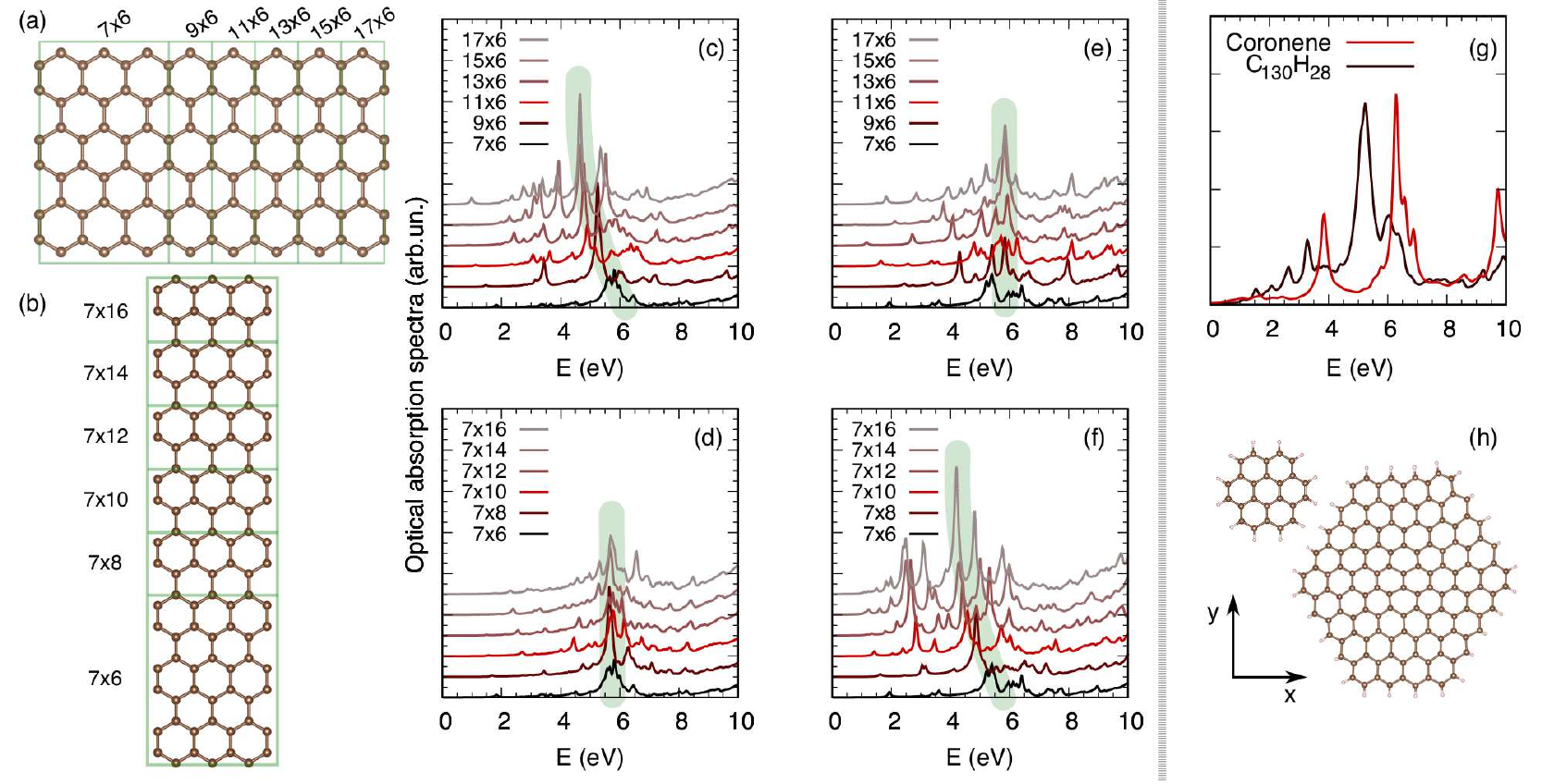}
\caption{\label{fig: figure4} Size and shape effect on the optical spectra. Panel (a) shows the structures of the rectangular flakes with constant armchair edge: 7$\times$6-RGNF is the smallest flake and the rest are constructed by increasing its size by two vertical dimer lines. Flakes with constant zigzag edge are shown in (b), starting from the same 7$\times$6-RGNF they grow by adding two horizontal zigzag lines. In both (a) and (b), carbon atoms are represented in brown and the passivating hydrogens are not drawn for the sake of clarity. (c-f) contain the rectangular flakes' in-plane optical spectra for different sizes and electric field directions (x-direction is parallel to the zigzag edge and y-direction is parallel to the armchair edge), obtained from TDDFT calculations. Panel (c) shows the x-component of the optical spectra for the flakes with constant armchair edge, panel (e) shows the y-component for the same flakes. (d) contains the x-component for the constant zigzag-edged flakes and finally (f) shows their y-component. As is seen in panel (c) and (f), an increase in length along the field leads to absorption at lower energies, whereas the constant edge-direction, (e) and (d), show no major changes in the absorption onset. Finally, (g) shows the normalized optical spectra for two hexagonal molecules, C$_{130}$H$_{28}$ and coronene, and their structures are presented in (h), including passivating hydrogen in white.}
\end{figure*}

Changing the shape and size of graphene nanoflakes allows, in addition to tailor the strength of the magnetic exchange coupling, to modify the optical spectrum. To systematically investigate the contribution of zigzag and armchair edges to the optical spectra, we construct a series of RGNF. The starting point is a 7$\times$6-RGNF which we extend in the x and y direction separately, as shown in figure \ref{fig: figure4}, to construct 11 different flakes. The procedure is motivated by the standard experimental bottom-up growth of graphene nanoribbons \cite{Cai2010}, by halting the process early, rectangular nanoflakes of specific dimensions can be synthesized with high accuracy. This series of RGNF have the same magnetic properties as the ones discussed for 17$\times$6-RGNF, but we choose to focus on the contribution of their size and shape to the optical properties and not on their magnetism. The total optical absorption spectra shown in panels (a) and (b) of the same figure show  the expected tendency to close the gap with increasing flake size. Furthermore, by growing the flakes in the armchair direction and keeping the zigzag edge, the first peak tends to appear at lower energies. 


In order to understand the optical spectra in depth, we plot their directional contributions in Figure \ref{fig: figure4}. The green trend line shows that with increasing flake size the main peak position shifts to lower energies only in the direction parallel to the applied electric field, while applying an electric field parallel to the edge that is kept constant shows no peak shift.

However, having zigzag edges alone is not a sufficient condition for magnetism, as it was pointed out in a previous publication \cite{MansillaWettstein2016}. Small hexagonal flakes with zigzag edges are not magnetic, and we report the optical spectra for coronene and its large analogue, C$_{130}$H$_{28}$. Their spectra and structures are shown in figure \ref{fig: figure4}, where we observe in (a) the presence of the coronene peaks in C$_{130}$H$_{28}$, although they appear at lower energies due to quantum confinement. Furthermore, the large flake has multiple peaks before the coronene peak. 

The comparison between coronene and C$_{130}$H$_{28}$ shows that if the shape and edge structure are kept the same, the position of the major absorption peak is determined by the number of atoms or flake size. However, the shape and edge structure play a very important role in the peak position, as it can be seen when comparing the rectangular and hexagonal flakes. If only the size is taken into account, one would expect the major peak in 7$\times$16- and 17$\times$6-RGNF to be at higher energies than the one in C$_{130}$H$_{28}$ as the latter one is larger. However, the rectangular shape and the presence of armchair edges shift the peaks of the RGNF to lower energies. Thus, it is evident from our results that the flake shape and edge structure have defining features in the optical spectra, and that the positions of the peaks can be tuned by varying their size.


\section{Conclusions} 
We have shown that the magnetic exchange coupling between the magnetic moments on the zigzag edge can be tailored by optimizing the shape and size of rectangular graphene nanoflakes. This in turns allows for optimizing the switching conditions between AFM and FM coupling, for example tuned to energy efficiency by countering stability to the magnitude of the external field. Additionally, we have shown that there are clear changes in the optical absorption coefficient between different magnetic orderings. The possibility to alter the optical absorption spectrum with the size and shape of the flakes offers another degree of versatility. The coupling between these phenomena allows energy efficient device design. For example, optical detection of small currents allows for a route to construct fast electrical failure detection.

\section{Acknowledgements}
RE and OG acknowledge financial support from the strategic research council (SSF) grant  ICA16-0037. BS and BP are grateful to Swedish Research Council for supporting project grant (2016-05366) and Swedish Research Links programme grant (2017-05447). RE, OG and BS acknowledge Swedish National Infrastructure for Computing (SNIC) for the allocation of computing time in Tetralith, NSC.  BS acknowledges financial support from Energymyndigheten (Diary no. 2018-004434 \& Project number 46621-1). BP and RS thank Department of Science and Technology of Govt. of India   for financial support (DST-PURSE) and C-DAC, Pune for partial computing support.



%

\end{document}